\documentclass[lettersize,journal,12pt]{IEEEtran}
\usepackage{amsmath,amsfonts}
\usepackage{algorithmic}
\usepackage{algorithm}
\usepackage{array}
\usepackage[font=normalsize]{caption}
\usepackage{textcomp}
\usepackage{stfloats}
\usepackage{url}
\usepackage{verbatim}
\usepackage{graphicx}
\usepackage{cite}
\usepackage{bm}
\usepackage{nicefrac}
\usepackage{lipsum}
\usepackage{relsize}
\usepackage{microtype}
\usepackage{multirow}
\usepackage{color}
\usepackage[export]{adjustbox}
\usepackage{subfigure}

\newcommand{\biographyspace}{\vspace{-13mm}}

\begin{document}
%\relscale{0.95}
%\linespread{0.95}

\title{\vspace{-4mm}Demonstrating Interoperable Channel State Feedback Compression with Machine Learning}

\author{Dani Korpi, Rachel Wang, Jerry Wang, Abdelrahman Ibrahim, Carl Nuzman, Runxin~Wang, Kursat Rasim Mestav, Dustin Zhang, Iraj Saniee, Shawn Winston, Gordana~Pavlovic, Wei Ding, William J. Hillery, Chenxi Hao, Ram Thirunagari, Jung~Chang, Jeehyun Kim, Bartek Kozicki, Dragan Samardzija, Taesang Yoo, Andreas~Maeder, Tingfang Ji and Harish Viswanathan \vspace{-12mm}
        % <-this % stops a space
\thanks{This work has been partly funded by the European Commission through the project Hexa-X-II (Grant Agreement no. 101095759).}
\thanks{Dani Korpi, Jerry Wang, Carl Nuzman, Kursat Rasim Mestav, Iraj Saniee, Gordana Pavlovic, Ram Thirunagari, Bartek Kozicki, Dragan Samardzija, and Harish Viswanathan are with Nokia Bell Labs, Rachel Wang, Abdelrahman Ibrahim, Runxin Wang, Dustin Zhang, Shawn Winston, Wei Ding, Jung Chang, Taesang Yoo, and Tingfang Ji are with Qualcomm Technologies, Chenxi Hao is with Qualcomm Wireless Communication Technologies, William J. Hillery, Jeehyun Kim, and Andreas Maeder are with Nokia Standards.}% <-this % stops a space
}


\maketitle

\begin{abstract}
Neural network-based compression and decompression of channel state feedback has been one of the most widely studied applications of machine learning (ML) in wireless networks. Various simulation-based studies have shown that ML-based feedback compression can result in reduced overhead and more accurate channel information. However, to the best of our knowledge, there are no real-life proofs of concepts demonstrating the benefits of ML-based channel feedback compression in a practical setting, where the user equipment (UE) and base station have no access to each others' ML models. In this paper, we present a novel approach for training interoperable compression and decompression ML models in a confidential manner, and demonstrate the accuracy of the ensuing models using prototype UEs and base stations. The performance of the ML-based channel feedback is measured both in terms of the accuracy of the reconstructed channel information and achieved downlink throughput gains when using the channel information for beamforming. The reported measurement results demonstrate that it is possible to develop an accurate ML-based channel feedback link without having to share ML models between device and network vendors. These results pave the way for a practical implementation of ML-based channel feedback in commercial 6G networks.
\end{abstract}

\section{Introduction} 

\IEEEPARstart{I}{n} recent years, there have been various proposals for how to apply artificial intelligence (AI) and machine learning (ML) in cellular networks \cite{Hoydis21,hexa2_d43}. The example use cases range from physical layer \cite{Honkala21} to higher level network management \cite{Morocho19}. It is even envisioned that 6G could be built to incorporate a completely AI-native air interface \cite{Hoydis21}. All of these applications are justified by the increased performance, accuracy, robustness, and flexibility provided by the AI/ML-based solutions.

One of the more established research topics for use of AI/ML in cellular networks is to apply it to the compression of channel state information (CSI) feedback between the user equipment (UE) and the base station (gNB) \cite{guo_2022}. {This use case has been the subject of an official study item in 3rd Generation Partnership Project (3GPP) Releases 18 and 19 \cite{3GPPTR38.843} and has been approved as a normative work item for Release 20. Therefore, ML-based CSI feedback compression is likely to see more widespread adoption in the 6G era.} In particular, AI/ML-based CSI feedback compression involves a two-sided model, where the encoder part of the model at the UE side compresses the channel information, while a compatible AI/ML-based decoder part of the model is used at the gNB side to decompress and reconstruct the channel feedback \cite{guo_2022,3GPPTR38.843}. It is envisioned that such a two-sided model framework developed for the ML-based CSI feedback could serve as a stepping stone for a completely AI-native air interface.

The main benefit of AI/ML-based CSI feedback compression is a more favorable trade-off between compression ratio and accuracy. This means that one can either use a higher compression ratio without reducing the accuracy, or a higher accuracy with the same compression ratio. Typically the CSI feedback is used for downlink (DL) beamforming, meaning that any improvement in the accuracy of the CSI feedback is likely to also improve the DL throughput. In other cases, it might be more preferable to reduce the overhead of the feedback, in which case one can utilize an encoder-decoder pair that achieves a higher compression ratio. This could be beneficial especially in high-order MIMO scenarios, where the volume of CSI feedback is high.

There is a wide body of existing literature on theoretical and simulation-based studies of ML-based CSI feedback compression \cite{Dreifuerst24a,guo_2022,Zimaglia20,Guo20a}, as well as some experimental works \cite{Cheng24,Shehzad24a}. Prior art on deep autoencoder architectures for CSI compression includes models such as \textit{CsiNet}, \textit{CsiNet-LSTM}, \textit{deepCMC}, \textit{PolarDenseNet}, and \textit{CsiTransformer} which leverage high-fidelity neural network autoencoder architectures based on convolutional or transformer models~\cite{guo_2022}. However, there are various practical challenges that are yet to be addressed. Especially, a crucial aspect of such two-sided ML models is to ensure that the multiple encoder models employed by the UEs are compatible and interoperable with the decoder model at the gNB. What is more, in order to protect trade secrets and confidential product implementations, it is often not desirable to disclose the exact model architecture or parameters to the other party. This means that the ML models must sustain a robust and accurate CSI feedback link under real propagation channels while being trained separately.

To this end, we propose a novel method for training interoperable encoder and decoder models in a confidential manner, while also showing that a single decoder model can interoperate with multiple encoder models. Moreover, we demonstrate with a proof-of-concept (PoC) implementation that ML-based CSI feedback compression works in the real world and outperforms a non-ML baseline. To the best of our knowledge, this is the world's first real-life demonstration of ML-based CSI feedback compression between a UE vendor and a network vendor, represented by Qualcomm and Nokia, respectively.

\section{CSI Feedback Compression in Cellular Networks}
\label{sec:background}

\begin{figure*}[!t]
\centering
\includegraphics[width=0.6\textwidth,trim=0 320 250 0, clip]{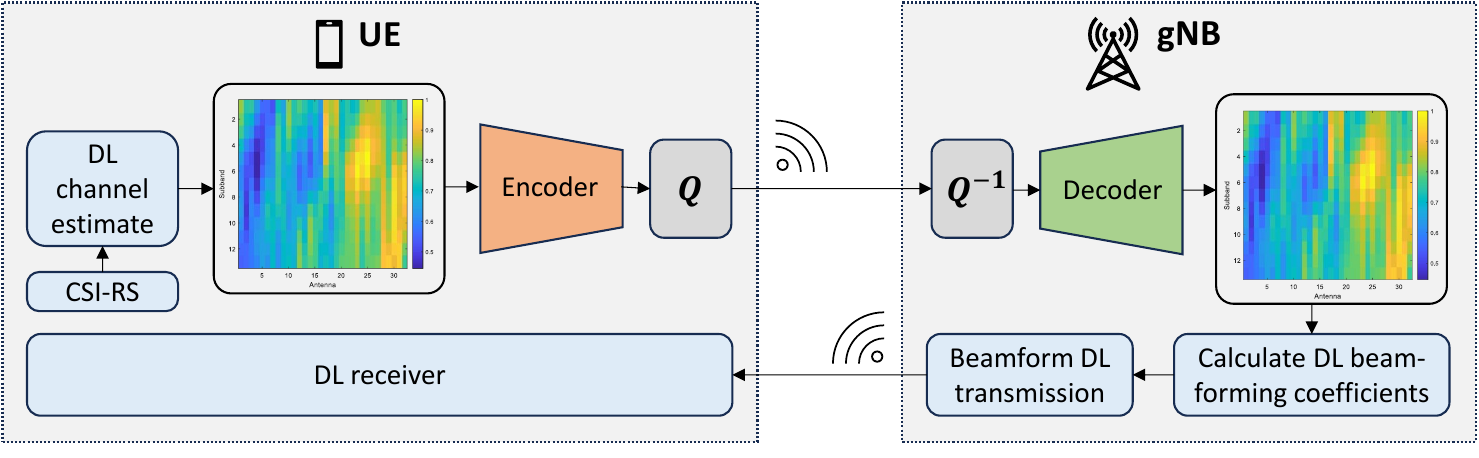}
\caption{The considered framework for ML-based CSI feedback compression.}
\label{ae-fig}
\end{figure*}

We consider a DL massive MIMO system between a single gNB and a single UE. The high-level system model is depicted in Fig.~\ref{ae-fig}, which illustrates the elements that are the focus of this work. Both the gNB and the UE have multiple antennas, and the system utilizes an orthogonal frequency-division multiplexing (OFDM) waveform. The objective of the gNB is to deliver a predefined number of spatial streams to the UE with the highest possible throughput. The number of spatial streams is often referred to as the number of MIMO layers.

In this work, we focus on optimizing this objective in terms of DL beamforming accuracy. Namely, to maximize DL system throughput, the gNB needs to calculate the beamforming coefficients, i.e., the precoder, based on accurate knowledge of the DL CSI. It is well known that, for a single-user MIMO, the optimal choice of precoder for a given number of MIMO layers is to use the strongest eigenvectors of the channel covariance matrix as the beamforming weights. This strategy maximizes the closed loop MIMO channel capacity.

In a time-division duplex (TDD) system with ideal uplink-downlink reciprocity, the CSI for calculating the beamforming weights is available at gNB via measurement of the uplink (UL) sounding signal. In a frequency-division duplex (FDD) system or a TDD system with non-ideal reciprocity, the CSI is acquired via UE report. More specifically, the UE first estimates the DL channel based on CSI reference signal (CSI-RS) transmitted by the gNB, as depicted in Fig.~\ref{ae-fig}. Then, the UE extracts the required information from the estimated CSI and reports it to the gNB. The information carried in this CSI feedback message is designed such that it allows the gNB to determine the precoding vectors.

There are two general directions of CSI reporting: \textit{explicit} CSI reporting and \textit{implicit} CSI reporting. The former reports the raw channel measurements, while the latter reports the precomputed precoding coefficients. {Especially in high-order MIMO scenarios, explicit channel reporting can be expected to consume more bandwidth as it requires the UE to report the complete channel data, as opposed to per-layer precoding coefficients reported in implicit CSI feedback.} In this work, we focus on implicit CSI, and aim at developing a practical ML-based method for efficient compression while maintaining good CSI accuracy. {However, it should be emphasized that the developed and presented methods are also in principle applicable to compressing explicit CSI, which is one of our potential future research directions. For now, more details on explicit CSI compression can be found, e.g., in \cite{Wen18a}.}

In the past two decades, various CSI codebooks have been developed for performing CSI compression in spatial, frequency and even in time domain~\cite{ts38214}. Referring to Fig.~\ref{ae-fig}, in this paper we assume that an ML-based CSI encoder is used at UE side to compress the precoders to a latent message. The latent message is then quantized (this operation is denoted by $Q$ in the figure) and the binary representation of the quantized latent message is delivered to the gNB, assuming error-free transmission. Note that the quantization (and the subsequent de-quantization) is done using a predefined codebook that is shared between gNB and UE.

In the gNB, the quantized latent message is first extracted from the binary representation via a de-quantizer, denoted by $Q^{-1}$ in Fig.~\ref{ae-fig}. Then, an ML-based CSI decoder is used for reconstructing the CSI feedback based on the quantized latent message. The reconstructed CSI feedback, which consists of the precoding coefficients, is then used for beamforming the consecutive DL transmissions. Note that it is assumed that a common precoder is used for all the subcarriers within the same sub-band, similar to legacy enhanced type II (eType-II)-based CSI reporting \cite{ts38214}. Ultimately, the accuracy of the CSI feedback is measured by the achieved DL throughput. However, since estimating the DL throughput requires comprehensive measurement or simulation campaigns, we use squared generalized cosine similarity (SGCS) between the original and decoded precoders as an additional measure of the CSI accuracy.

\section{Training Interoperable ML Models}
\label{sec:interoperability}

\begin{figure*}%
\centering
\subfigure[]{%
\label{fig:first}%
\includegraphics[width=0.3\textwidth,trim=0 380 680 0, clip,valign=t]{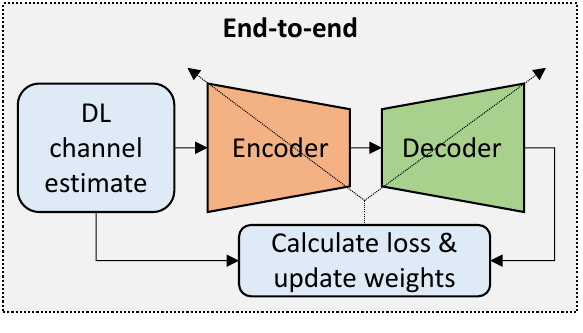}}
\quad
\subfigure[]{%
\label{fig:second}%
\includegraphics[width=0.3\textwidth,trim=0 240 670 0, clip,valign=t]{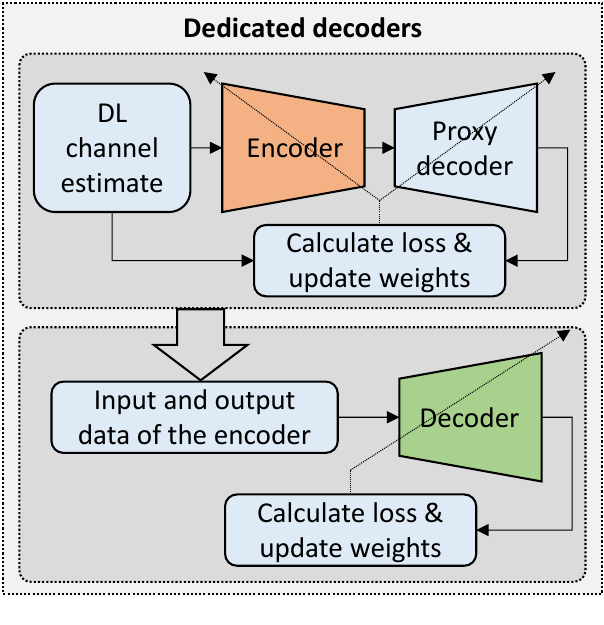}}
\quad
\subfigure[]{%
\label{fig:third}%
\includegraphics[width=0.3\textwidth,trim=0 230 660 0, clip,valign=t]{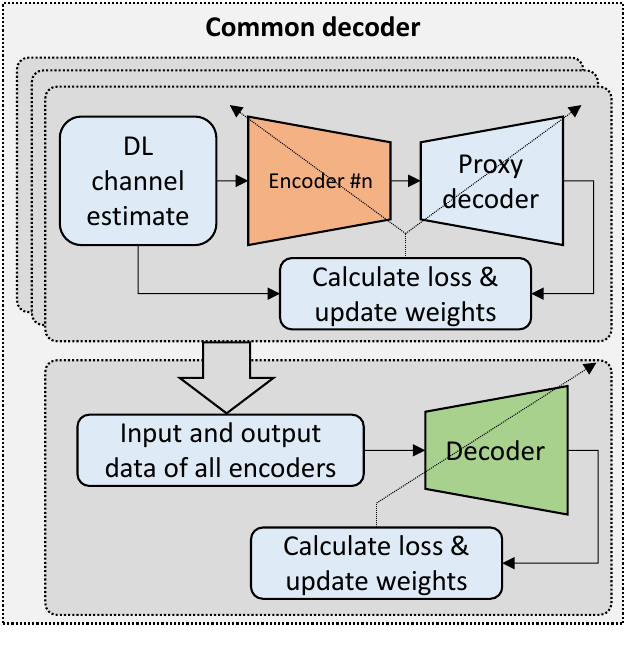}}
\caption{The different approaches for training the encoder and decoder models: (a) joint end-to-end training, (b) sequential training with a dedicated decoder, and (c) sequential training with a common decoder for multiple encoder models.}
\vspace{-3mm}
\label{training-fig}
\end{figure*}

As discussed above, our PoC system relies on implicit CSI feedback, which consists of the eigenvectors of the estimated transmit covariance matrices, i.e., the precoding coefficients, aggregated across frequency-domain sub-bands. {In our testbed, the dimensions of the matrices are 8 (Tx antennas) by 14 (sub-bands) by 2 (real/imaginary) by 16 bits which are measured for each MIMO layer and across the whole bandwidth. With 70 sub-bands and 4 layers, this translates to 71680 bits which are reported every 10 ms.\footnote{It should be noted that in practice there are channel measurements for 68 physical sub-bands, with two sub-bands being copied before the ML encoder to have 70 effective sub-bands. This ensures that the total bandwidth is divisible into blocks of  14 sub-bands, which is the processing size of the encoder and decoder models.} In our PoC, we assume a fixed compression ratio of 28x and aim to compress this array down to 2560 bits / 10 ms using deep autoencoder models. The encoder model generates a 128-bit output for each layer and each block of 14 sub-bands, resulting in a total of 20 inferences.} Naturally, the lower the reconstruction error with the same compression ratio, the better the encoder-decoder model. The different approaches for training the compression and decompression ML models are depicted in Fig.~\ref{training-fig} and described in more detail below. {For more insights on extending ML-based CSI compression to support multiple compression ratios, we refer the reader to prior literature \cite{Guo20a}.}

\subsection{Encoder and Decoder Models}
\label{sec:models}

In this work, the ML models utilized for compression and decompression of the CSI feedback are based on architectures adopted from image processing domain. In particular, two ML encoder models are considered, one of them based on convolutional neural network (CNN) architecture and another based on transformer (TF) architecture. The TF encoder is a variation of a vision transformer (ViT) architecture adopted from \cite{Dosovitskiy21a}, where it was shown to achieve good image classification accuracy compared to CNN-based state-of-the-art benchmarks. As for the ML decoder model, it also follows a TF architecture based on an appropriately modified ViT model \cite{Dosovitskiy21a,ChenEtAl25}. {To achieve the desired compression ratio, the output of all considered encoder models has a latent dimension of 64 with 2 bits per dimension, i.e., 128 bits in total. This represents the compressed CSI data of one MIMO layer over 14 sub-bands.}

The most straightforward training approach for this type of an ML task is the joint end-to-end training depicted in Fig.~\ref{training-fig}~(a), where the encoder and decoder models must be fully exposed \cite{3GPPTR38.843}. As discussed earlier, such an approach is problematic due to the issue of multi-vendor interoperability and lack of confidentiality.

\subsection{Sequential Training}
\label{sec:dedicated_dec}

To address these shortcomings, in this PoC system we use a UE-First separate sequential training framework. Focusing first on the approach where the gNB uses a dedicated decoder for each UE encoder, illustrated in Fig.~\ref{training-fig}~(b), the first step is for the UE vendor to train a deep autoencoder containing both an encoder and a decoder, with quantizer and inverse quantizer between the two (the quantizer is omitted from Fig.~\ref{training-fig} for clarity). The resulting encoder is to be used at the UE, while the decoder, referred to as a proxy decoder, is only used for training purposes by the UE. The second step, after the UE autoencoder has been fully trained, is for the UE vendor to create a large enough data set (i.e., 100~000s of samples) of CSI matrices and the corresponding dequantized latent vectors obtained by passing the CSI matrices through the sequence of UE encoder, quantizer, and inverse quantizer.
 
The resulting data set is shared with the gNB vendor and it provides an implicit description of the UE encoder; no other information about the UE autoencoder architecture or parameters needs to be shared. The final step is for the gNB vendor to train its decoder using the dequantized latent vectors as input and corresponding CSI matrices as output. The loss function the gNB uses for training its decoder is naturally the error between the CSI matrix from the UE training samples and the resulting output from the decoder at the gNB, measured by SGCS. Once the decoder has been successfully so trained, and the UE vendor has provided a codebook for the dequantization function, the resulting combination of the UE's encoder, quantizer, inverse quantizer, and gNB's decoder is ready to be used operationally, as demonstrated in Section~\ref{sec:results}.

\subsection{Multi-vendor Training}
\label{ssec:multivendor}

One question being debated in 3GPP is whether, in a UE-first separate training context, it is practical for different UE vendors to develop different proprietary encoders. In principle, a gNB vendor could support this paradigm by training a dedicated decoder for each encoder independently on datasets provided by each UE vendor, following the sequential approach of the previous subsection. However, storing and quickly accessing a separate set of trained decoder weights for each UE vendor would incur a significant implementation cost.

To overcome this obstacle, we developed a novel approach of training a decoder with a single set of weights capable of decoding latent vectors from multiple UE encoders. In this approach, depicted in Fig.~\ref{training-fig}~(c), we have a number of different UE vendors, where each vendor provides a dataset as described in the previous subsection. All datasets are combined, and the decoder input vector is increased by one dimension to an extended vector that includes an index defining which dataset, i.e., UE encoder, the latent vector came from. The decoder is then trained to map these extended vectors to corresponding CSI matrices. This approach is more scalable than having a library of dedicated decoders for each UE vendor, and in Section~\ref{sec:results} we also demonstrate that it can achieve CSI reconstruction accuracy comparable to the other schemes.

{\subsection{Additional Remarks about Model Training}}

{As an additional remark on the proposed scheme, it should be noted that it is also possible to utilize the sequential training approach in reverse order. In that case, the gNB vendor trains the decoder model first, using a proxy encoder. Once the gNB decoder has been fully trained, a data set consisting of decoder inputs and corresponding CSI matrices is produced, taking into account the appropriate quantization steps. The UE vendor can then train a compatible encoder model using these as target outputs and inputs, respectively. Although this approach should yield fundamentally similar compression performance as the UE-First approach, it allows for the encoder model to be retrained without having to necessarily retrain the decoder model. This introduces some autonomy for the UEs, which can be beneficial in dynamic channel environments. However, for brevity and clarity, the focus of this article is solely on UE-First training.}

{Secondly, we wish to emphasize the fact that in most cases the training data consists of measured channel responses between the gNB and the UE, which contain some level of information about the environment around the UE. Therefore, the privacy aspects of the data should be carefully considered in any commercial deployment. In our work, the training data was measured using our own prototype and test hardware, which means that no user data was used. Another way to avoid all possible privacy issues is to train the models using synthetic channel data.}

\section{Proof of Concept Implementation}
\label{sec:results}

\begin{figure*}[!t]
\centering
\includegraphics[width=\textwidth,trim=0 180 160 0, clip]{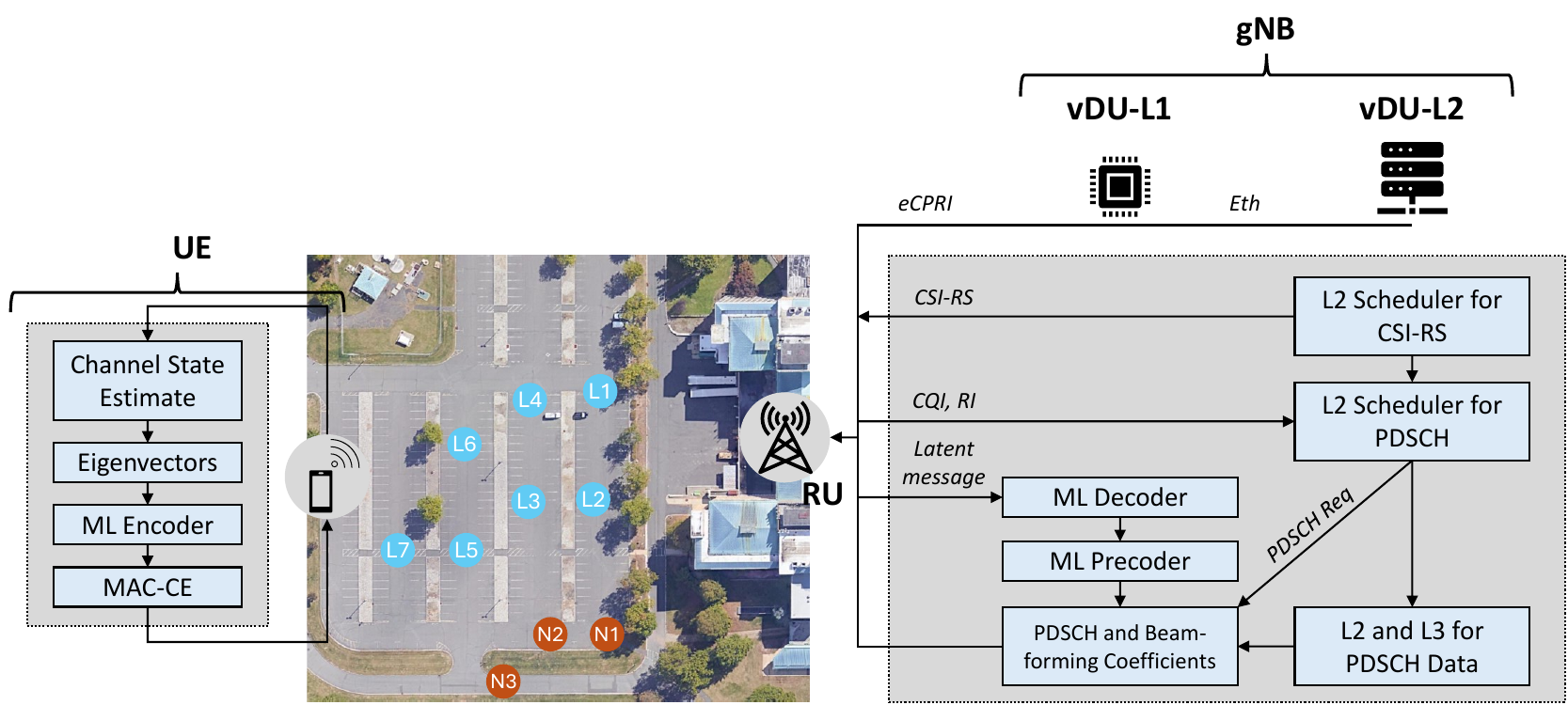}
\caption{A block diagram of the implemented hardware testbed and the over-the-air measurement locations. Map data: \textcopyright 2025 Google, Airbus, Maxar Technologies.}
\label{fig:testbed}
\end{figure*}

In order to investigate the different training approaches, as well as to perform further experimental evaluations of ML-based CSI feedback compression, we developed a hardware PoC testbed. The testbed is illustrated in Fig~\ref{fig:testbed}, and described in more detail below.

\subsection{Hardware Setup}

In the hardware setup, the mobile terminal is a Qualcomm development UE with four antennas and it includes firmware that supports ML-based CSI feedback compression. The Nokia Bell Labs experimental radio unit (RU) consists of an eight-antenna linear array with single polarization, enabling horizontal-plane beamforming with up to four MIMO layers. The gNB consists of virtualized distributed unit (vDU) Layer~1 (L1) and vDU Layer~2 (L2) servers. L1-high real-time processing is implemented on a programmable Intel CPU and Nvidia GPU platform. {The GPU model used in the prototype gNB is Nvidia GeForce RTX 3080.}

The L2 vDU scheduler sets the 10 ms CSI-RS periodicity. Upon receiving the CSI-RS signal, the UE performs channel estimation for all sub-bands, transmit antenna ports, and receive antennas, and calculates the precoding coefficients. {The resulting CSI is then fed through the ML encoder to obtain the CSI feedback message that is compressed by a ratio of 28x, as discussed earlier.} The ML unit can be configured with either a CNN or a TF model.

The output of the ML encoder consists of 128 bits that represent the precoding coefficients of a single layer over 14 sub-bands, which together span a bandwidth of 56 physical resource blocks (PRBs), or approximately 20~MHz. Altogether, with 4 layers and a bandwidth of 100~MHz, the total number of ML encoder inferences is 20, resulting in a 2560-bit feedback message. The encoded CSI is packed into a medium access control (MAC) - control element (CE) message transmitted in the uplink. Channel quality index (CQI), rank indicator (RI), and encoder ML model ID are also inserted in the MAC-CE packet.

The vDU L1 server at the network side receives the MAC-CE message, extracts the CQI and RI fields, and passes them to L2 for physical downlink shared channel (PDSCH) scheduling. L1 also extracts the CSI feedback message and performs the ML decoding based on the encoder ML model ID indicated by the UE, producing the corresponding decompressed CSI. The gNB can use either a common or dedicated model, as described in Section~\ref{sec:interoperability}. {With the utilized GPU model, the inference time of the ML-based decoder was observed to be in the order of 0.32~ms (across 4 MIMO layers and 100~MHz of bandwidth). In the prototype system, such inference latency was deemed sufficient, although a more favorable trade-off between memory/energy consumption and inference latency can likely be obtained with custom ML inference hardware.}

After the ML decoder, the decompressed CSI is fed into the ML precoder unit for generating the ML precoding coefficients. In particular, even though the original UE-generated eigenvectors are mutually orthogonal, re-orthogonalization must be performed in the precoder unit to mitigate cross-layer interference caused by the reconstruction errors. Finally, the PDSCH data and ML precoding coefficients are transported over an eCPRI link to the RU. Within the RU, beamforming is applied, and the corresponding signal is fed to the antennas. In the current system, spatial precoding is applied only on the PDSCH data, excluding the CSI-RS signal. Consequently, the UE channel estimation is based on non-coded signal.

\subsection{Performance Results of Sequential and Multi-vendor Training}

First, the encoder/decoder pairs using the three different training approaches described in Section~\ref{sec:interoperability} are tested using channel feedback data extracted from the hardware testbed. In the first step, the UE partner, represented by Qualcomm, trains two end-to-end autoencoder models. Both models have the same TF-based decoder architecture, but one uses a CNN-based encoder while the other uses a TF-based encoder. The SGCS performance of each of these two autoencoders is tested using a test dataset separate from the datasets used for training. The results are reported in the second column of Table~\ref{table:seqPerf} for both encoders. In the table, we report the average SGCS performance separately for each eigenvector rank, as well as the average over all ranks. The transformer architecture performs slightly better than the CNN architecture, but both give very accurate results. It can also be seen in Table~\ref{table:seqPerf} that the SGCS performance decreases with increasing eigenvector rank. This is because the eigenvectors corresponding to the stronger layers, which are dominated by simple line-of-sight (LOS) propagation, are easier to compress than the eigenvectors of the weaker layers that are associated with more complex multi-path propagation modes. It should be noted that this phenomenon is not dependent on the compression method.

Then, in order to test the sequential training approach, the UE partner shared a dataset (as described in Section~\ref{sec:dedicated_dec}) for each of the two encoders with the gNB partner, represented by Nokia. The gNB partner then independently trained a TF-based decoder architecture on each dataset, obtaining a dedicated decoder for each encoder type. After this, the reconstruction accuracy of the dedicated decoder for each of the two encoder types was tested. The results are shown in the third column of Table~\ref{table:seqPerf}.  As the table shows, the SGCS obtained using sequential training is virtually identical to the performance originally obtained by the UE vendor for the corresponding end-to-end autoencoders. 

\begin{table}[t]
\setlength{\tabcolsep}{5pt}
\centering
\caption{{Average CSI reconstruction accuracy (SGCS) under the different training methods, when the compression ratio is fixed to 28x.}\label{table:seqPerf}}
\label{tab:accuracy}
\begin{tabular}{|c|c|c|c|}
\hline
\textbf{Encoder} & \textbf{End-to-end} & \textbf{Dedicated}  & \textbf{Common}  \\
\textbf{(Type)} & \textbf{(UE)} & \textbf{Decoders (gNB)}  & \textbf{Decoder (gNB)} \\
\hline
CNN Eig 1  & 0.973 & 0.976 & 0.969  \\
CNN Eig 2 & 0.872 & 0.883  & 0.871  \\
CNN Eig 3 & 0.759 & 0.756  & 0.748  \\
CNN Eig 4 & 0.655 & 0.643  & 0.638  \\
\hline
\textbf{Average} & 0.815 & 0.814  & 0.806  \\
\hline
TF Eig 1 & 0.977 & 0.975  & 0.973  \\
TF Eig 2 & 0.886 & 0.890  & 0.890  \\
TF Eig 3 & 0.781 & 0.771  & 0.772  \\
TF Eig 4 & 0.683 & 0.665  & 0.668  \\
\hline
\textbf{Average} & 0.823 & 0.820  & 0.816  \\
\hline
\end{tabular}
\end{table}

Finally, to test the performance of multi-vendor training, the gNB partner trained a single, common decoder for both encoders using a combined dataset, with an extended input vector as described in Section~\ref{ssec:multivendor}. The results are shown in the fourth column of Table~\ref{table:seqPerf}. The performance of the common decoder is only marginally lower than the performance obtained with dedicated decoders or with the original proxy decoders trained end-to-end by the UE vendor. These results demonstrate that full interoperability can be achieved 1) without needing to share information about encoder or decoder architectures or parameters between the vendors and 2) using a carefully trained common decoder that operates with different proprietary encoders with good reconstruction accuracy. These aspects can greatly simplify the gNB decoder implementation when working with multiple UE vendors in the same network.

\subsection{Real-Time Performance Evaluations}

\begin{table*}[ht]
\renewcommand{\arraystretch}{1.1}
\setlength{\tabcolsep}{5pt}
\centering
\caption{DL throughput gain evaluation of the ML-based CSI feedback compression PoC.\vspace{-3mm}}
\label{table:tput_gain}
\begin{tabular}{|l|l|l|l|l|l|l|l|l|l|l|l|}
\hline
\multirow{2}{*}{\textbf{\parbox{1.7cm}{Test location (see Fig.~\ref{fig:testbed})}}} &  \multicolumn{7}{c|}{\textbf{Line of sight (LOS)}} & \multicolumn{3}{c|}{\textbf{Non line of sight (NLOS)}} & \multirow{2}{*}{\textbf{Average}}   \\
\cline{2-11}
& \textit{\textbf{L1}} & \textit{\textbf{L2}} & \textit{\textbf{L3}} & \textit{\textbf{L4}} & \textit{\textbf{L5}} & \textit{\textbf{L6}} & \textit{\textbf{L7}} & \textit{\textbf{N1}} & \textit{\textbf{N2}} & \textit{\textbf{N3}} & \\
\hline
\textit{\textbf{Indoor model}} & 50.30\% & 95.60\% & 61.90\% & 76.60\% & 74.30\% & 39.10\% & 70.50\% & 39.60\% & 96.10\% & 61.10\% & 66.51\% \\
\hline
\textit{\textbf{Outdoor model}} & 41.60\% & 83.10\% & 65.30\% & 57.90\% & 69.00\% & 30.40\% & 67.90\% & 47.50\% & 100.40\% & 54.90\% & 61.80\% \\
\hline
\textit{\textbf{Mixed model}} & 50.30\% & 94.70\% & 65.70\% & 77.30\% & 86.90\% & 39.10\% & 72.10\% & 28.40\% & 96.40\% & 64.20\% & 67.51\%\\
\hline
\end{tabular}
\end{table*}

\begin{figure}[!t]
\centering
\includegraphics[width=\columnwidth,trim=110 286 130 297, clip]{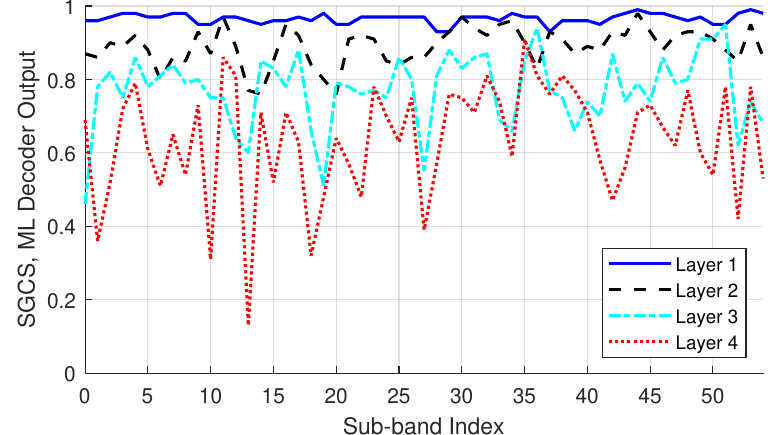}
\caption{Over-the-air SGCS results for the ML decoder output.}
\label{fig:sgcs_results}
\end{figure}

Finally, in order to evaluate the real-time performance gain of the considered ML-based CSI feedback compression scheme over the conventional approach, we conducted a number of over-the-air (OTA) evaluations. Specifically, the tests were conducted at the Nokia Bell Labs campus in Murray Hill, New Jersey. The UE device was placed in a vehicle, while the RU was installed on the fifth floor window overlooking a large parking lot. Various stationary and low-mobility scenarios were considered, the approximate measurement positions being indicated in Fig.~\ref{fig:testbed}. All of the measurement results were obtained with sequentially learned encoder and decoder models, as described in Section~\ref{sec:interoperability}.

As a first illustration, Fig.~\ref{fig:sgcs_results} shows the SGCS results based on a snapshot of the real-time OTA data, calculated from the ML decoder output. The SGCS is compared with the original encoder input eigenvectors (before the ML compression). The results show that the SGCS values of the strongest layers 1 and 2 are consistently high over all sub-bands, indicating robust interoperability of the proposed ML-based CSI feedback compression framework. There is more fluctuation in the weaker layers due to the fact that their respective eigenvectors are associated with more complex multi-path propagation modes, making them fundamentally more challenging to compress.

Next, the throughput of the developed PoC system is evaluated in the ten different OTA environments marked in Fig.~\ref{fig:testbed}, seven with LOS propagation and three with non-line-of-sight (NLOS) conditions. The results are summarized in Table~\ref{table:tput_gain}, where we list the mean DL throughput gain of the ML-based CSI feedback compression approach against conventional Type I CSI feedback in the OTA tests. At each location, we have measured the performance of multiple ML encoder-decoder pairs, which have been trained using the proposed sequential approach with one of three datasets: (i) indoor channel dataset, (ii) outdoor channel dataset, and (iii) a mixture of indoor and outdoor channel data. Note that in Table~\ref{table:tput_gain} we report only the performance corresponding to the TF-based ML encoder, for brevity.

Across all the different measurement locations, significant DL throughput gain is observed. The gain is ranging between 30\% and 100\%, with an average gain in the order of 60\%. As expected, the models trained with a mixture of indoor and outdoor data achieve the highest performance. However, interestingly, the indoor-trained model outperforms the outdoor-trained model. Our reasoning for this is that since the outdoor channel training data was measured under relatively high signal-to-noise ratios, the resulting ML models are less capable of dealing with the noisier channel estimates in the OTA measurement scenarios. {Nevertheless, all tested models achieve significant performance gains across all the different conditions, even when there is a mismatch between the training data and the deployment environment. This demonstrates the robustness and accuracy of the considered training approaches and ML model architectures. Moreover, the results reported in Table~\ref{table:tput_gain} also indicate that the trained models have generalized reasonably well since similar performance is observed with three different training data sets.}

It should be noted, however, that the Type I baseline utilizes wideband precoding, unlike the evaluated ML-based CSI feedback compression scheme which is sub-band precoded. This might somewhat inflate the measured throughput gains, although we do not consider it to be a major factor. Based on our assessment, the throughput gains can be mostly attributed to the improved compression performance of the utilized ML models.

\section{Key Findings and Future Directions}
\label{sec:conc}

Using ML to enhance the spectral efficiency of the next generation cellular networks is gaining significant traction, especially in the context of 6G standardization. To this end, it is important to evaluate the practical feasibility of the various ML-based algorithms. In this article, we presented and discussed practical methods for training interoperable ML models for CSI compression and decompression. A key requirement is that the UE and network vendors should not have to expose their ML models. Consequently, our proposed approach relies on the UE vendors sharing training data sets with the network vendor. This makes it possible for the network vendor to train an interoperable decoder model that can decompress the CSI feedback from multiple encoders without having access to the UE-side ML models.

To evaluate the real-world performance of the ensuing encoder and decoder models, we built a proof of concept consisting of UE and gNB prototypes and conducted over the air measurements using the ML models. Through these experiments we (i) confirmed interoperability between independently developed ML encoders and decoders under realistic conditions, (ii) demonstrated that a single gNB decoder can support multiple UE encoders, and (iii) showed improvements in the DL throughput.

{While these findings provide clear evidence of the real-world benefits of ML-based compression methods, industry-wide co-operation is needed to introduce commercially viable ML-based CSI feedback compression into the 3GPP specifications for 6G. Building on the AI/ML features standardized in 3GPP Release 19, implementation of ML-based CSI feedback compression is likely to require the following aspects:
\begin{itemize}
\item Means for ensuring compatibility between encoder and decoder, e.g., using an index-based approach for encoder/decoder pairing similar to what has been utilized in this work.
\item Coordination of training between the network and UE, e.g., via exchange of data sets as done in the proposed approach.
\item Means for data collection for training, where the data consists of ground truth CSI. This could require enhancement of traditional codebooks.
\item Means for monitoring the performance of the models and taking action if performance is not adequate.
\end{itemize}
These aspects have been investigated in a study item in Releases 18 and 19~\cite{3GPPTR38.843}, and recently 3GPP has approved a normative work item for Release 20, which will lead to the standardization of ML-based CSI feedback compression. This might also pave the way for standardization of other AI/ML use cases involving two-sided models, such as ML-based constellations and receivers \cite{Hoydis21}, and will therefore be an important milestone for the wider adoption of AI/ML-based methods in radio networks.}

{Finally, it should be noted there are still various AI and model-specific aspects requiring further research. For instance, the feasibility of interoperability should be explored across a wider range of model architectures, going beyond the transformer and CNN-based encoder models addressed in this article. Additionally, further work is needed in verifying the interoperability and practical feasibility of the more advanced CSI compression schemes, such as the ones performing prediction in addition to compression.}

\bibliographystyle{IEEEtran}

\bibliography{references}

\section{Biographies}
\vspace{-15mm}
\begin{IEEEbiographynophoto}{Dani Korpi}
is a Senior Specialist with Nokia Bell Labs, Espoo, Finland.
\end{IEEEbiographynophoto}

\biographyspace

\begin{IEEEbiographynophoto}{Rachel Wang}
is a Principal Engineer with Qualcomm Technologies, Inc., San Diego, CA, USA.
\end{IEEEbiographynophoto}

\biographyspace

\begin{IEEEbiographynophoto}{Jerry Wang}
is a Software Architect with Nokia Bell Labs, Murray Hill, NJ, USA.
\end{IEEEbiographynophoto}

\biographyspace

\begin{IEEEbiographynophoto}{Abdelrahman Ibrahim}
is a Staff engineer with Qualcomm Technologies, Inc., San Diego, CA, USA.
\end{IEEEbiographynophoto}

\biographyspace

\begin{IEEEbiographynophoto}{Carl Nuzman}
is a Research Manager with Nokia Bell Labs, Murray Hill, NJ, USA.
\end{IEEEbiographynophoto}

\biographyspace

\begin{IEEEbiographynophoto}{Runxin Wang}
is a Staff Engineer with Qualcomm Technologies, Inc., San Diego, CA, USA.
\end{IEEEbiographynophoto}

\biographyspace

\begin{IEEEbiographynophoto}{Kursat Rasim Mestav}
is a Machine Learning Algorithms Researcher with Nokia Bell Labs, Murray Hill, NJ, USA.
\end{IEEEbiographynophoto}

\biographyspace

\begin{IEEEbiographynophoto}{Dustin Zhang}
is a Principal Engineer/Manager with Qualcomm Technologies, Inc., San Diego, CA, USA.
\end{IEEEbiographynophoto}

\biographyspace

\begin{IEEEbiographynophoto}{Iraj Saniee}
is a Group Leader in the AI Lab, Nokia Bell Labs, Murray Hill, NJ, USA.
\end{IEEEbiographynophoto}

\biographyspace

\begin{IEEEbiographynophoto}{Shawn Winston}
is a Staff Engineer with Qualcomm Technologies, Inc., San Diego, CA, USA.
\end{IEEEbiographynophoto}

\biographyspace

\begin{IEEEbiographynophoto}{Gordana Pavlovic}
is a Fixed Networks Researcher with Nokia Bell Labs, Murray Hill, NJ, USA.
\end{IEEEbiographynophoto}

\biographyspace

\begin{IEEEbiographynophoto}{Wei Ding}
is a Principal Engineer/Manager with Qualcomm Technologies, Inc., San Diego, CA, USA.
\end{IEEEbiographynophoto}

\biographyspace

\begin{IEEEbiographynophoto}{William Hillery}
is a Senior Staff Research Specialist with Nokia of America, Naperville, Illinois.
\end{IEEEbiographynophoto}

\biographyspace

\begin{IEEEbiographynophoto}{Chenxi Hao}
is a Staff Engineer with Qualcomm Wireless Communication Technologies (China) Limited, Beijing, China.
\end{IEEEbiographynophoto}

\biographyspace

\begin{IEEEbiographynophoto}{Ram Thirunagari}
is a Software Developer with Nokia Bell Labs, Murray Hill, NJ, USA.
\end{IEEEbiographynophoto}

\biographyspace

\begin{IEEEbiographynophoto}{Jung Chang}
is a Senior Engineer with Qualcomm Technologies, Inc., San Diego, CA, USA.
\end{IEEEbiographynophoto}

\biographyspace

\begin{IEEEbiographynophoto}{Jeehyun Kim}
is a Senior Staff Research Specialist with Nokia Standards, Munich, Germany.
\end{IEEEbiographynophoto}

\biographyspace

\begin{IEEEbiographynophoto}{Bartek Kozicki}
is a Department Head with Nokia Bell Labs, Antwerp, Belgium.
\end{IEEEbiographynophoto}

\biographyspace

\begin{IEEEbiographynophoto}{Dragan Samardzija}
is a research lab leader with Nokia Bell Labs, Murray Hill, NJ, USA.
\end{IEEEbiographynophoto}

\biographyspace

\begin{IEEEbiographynophoto}{Taesang Yoo}
is a Senior Director of Technology with Qualcomm Technologies, Inc., San Diego, CA, USA.
\end{IEEEbiographynophoto}

\biographyspace

\begin{IEEEbiographynophoto}{Andreas  Maeder}
is Head of AI/ML with Nokia Standards, Munich, Germany.
\end{IEEEbiographynophoto}

\biographyspace

\begin{IEEEbiographynophoto}{Tingfang Ji}
is a Vice President of Engineering with Qualcomm Technologies, Inc., San Diego, CA, USA.
\end{IEEEbiographynophoto}

\biographyspace

\begin{IEEEbiographynophoto}{Harish Viswanathan}
is Head of Radio Systems Research Lab at Nokia Bell Labs, Murray Hill, NJ, USA.
\end{IEEEbiographynophoto}

\vfill

\end{document}